# Computing Entity Semantic Similarity by Features Ranking

Livia Ruback[1], Claudio Lucchese[23], Alexander Arturo Mera Caraballo[1],
Grettel Monteagudo García[1], Marco Antonio Casanova[1], Chiara Renso[2]

[1] Pontifical Catholic University of Rio de Janeiro, RJ, Brazil

[2] HPC Lab, ISTI-CNR, Pisa, Italy

[3] Ca' Foscari University of Venice, Italy

```
              liviaruback@gmail.com,
{lrodrigues, acaraballo, ggarcia, casanova}@inf.puc-rio.br,
       {claudio.lucchese, chiara.renso}@isti.cnr.it
```

***Abstract.*** *This article presents a novel approach to estimate semantic entity similarity using entity features available as Linked Data. The key idea is to exploit ranked lists of features, extracted from Linked Data sources, as a representation of the entities to be compared. The similarity between two entities is then estimated by comparing their ranked lists of features. The article describes experiments with museum data from DBpedia, with datasets from a LOD catalog, and with computer science conferences from the DBLP repository. The experiments demonstrate that entity similarity, computed using ranked lists of features, achieves better accuracy than state-of-the-art measures.*

## 1 Introduction

The emergence of the Linked Data initiative and the publication of several interlinked reusable data graphs representing entities opened the field of *Semantic Relatedness and Semantic Similarity* to measure the semantic relations between Linked Data resources, or *entities*.

Semantic similarity has the potential to be applied across many different domains: in the academic field, for example, it can be used to compare universities, departments, researchers or their works; in the tourism domain, to compare attractions, such as the museums visited during a trip. The computation of semantic similarity in the tourism, for example, provides rich datasets as input of data analysis to better understand the tourist behavior and find patterns and similarities between travelers to be exploited, for example, in recommendation systems.

This article addresses the problem of estimating entity similarity using entity features available as Linked Data. The key idea is to represent each entity by a list of features, extracted from Linked Data sources and ranked according to a relevance criterion. The similarity between two entities is then estimated by comparing their ranked lists of features, using rank correlation metrics to generate the similarity score. The exact features to be adopted and the relevance criteria are domain-dependent. The underlying hypothesis is that, by ranking the features based on their relevance, the accuracy of the similarity estimation is improved.

We assess the proposed approach to estimate entity similarity describing experiments in three domains: (i) museum descriptions found in DBpedia; (ii) datasets descriptions found in a Linked Open Data catalog; and (iii) computer science conferences available in the Linked Data version of DBLP[1]. In each of these domains, the experiments show that high-quality features are, respectively: the art movements of the artworks in a museum; the Wikipedia top-level categories that describe a dataset; and keywords extracted from computer science conference publications. The experiments results show that the proposed approach outperforms strong baselines in these domains in identifying the relevant features and computing the entity similarity.

In summary, this work makes the following contributions:

1. We *generalize* an approach to estimate entity similarity using ranked features extracted from Linked Data sources, called SELEcTOR, proposed in our previous work only for the museum domain [1];
2. We *prove the generality and domain-independent* characteristics of SELEcTOR, on instantiating the framework on two more domains: on comparing LOD datasets and computer science conferences.

In other words, this work presents as theoretical contribution a consolidated *methodology* to estimate entity similarity using ranked features extracted from Linked Data sources. The methodology is instantiated in three different domains with the description of extensive experiments to assess the proposed approach. It is important to mention that, thanks to its domain-independent nature, our approach can be applied to any domain. However, the insights we present here may profitably guide other experiments on the same domains – for instance, on choosing the best features to describe museums, datasets, and conferences, which represents a practical implication of our work.

The remainder of this article is structured as follows. Section 2 summarizes related work. Section 3 introduces preliminary concepts and algorithms used throughout the article. Section 4 describes the SELEcTOR framework, Section 5 discusses three experiments on different domains to assess the approach proposed to compare Linked Data entities. Finally, Section 6 reports the conclusions and future work.

## 2 Related Work

We divide related work into three groups, two of them focusing on *comparing* Linked Data Entities and a group focusing on *ranking* Linked Data entities: (i) Comparing Linked Data Entities with Graph-based and Wikipedia structure-based approaches; (ii) Comparing Linked Data Entities with ontology and Wikipedia structure-based approaches and (iii) Ranking Linked Data entities.

**Comparing Linked Data Entities with Graph-based and Wikipedia structure-based**

Several approaches profitably exploit the graph-based nature of the RDF structure and the concepts and relationship expressed in Wikipedia to compare or rank Linked Data Entities.

A known measure to compare Linked Data Entities is the *Wikipedia Link-based Measure* (WLM) [2], detailed in Section 3. It computes the relatedness between two Wikipedia articles a and b considering the links to a and b and the entire Wikipedia link structure.

---

[1] https://dblp.uni-trier.de

The *Semantic Connectivity Score* (SCS), proposed by Nunes et al. [3], computes the similarity between a pair of entities, based on the Katz score [4] and is detailed in Section 5. The measure computes the number of paths between the entities and penalizes longer paths.

Traverso et al. [5] proposed GADES, a graph-based similarity measure combining semantic aspects as neighbors, class hierarchies and node degrees and empirically evaluated the measure's accuracy in proteins and news domains. Ganggao et al. [6] proposed a method for measuring the semantic similarity between concepts in Knowledge Graphs (KGs), on combining the structure of the semantic network between concepts with their information content. Another related work in this group uses the museum's domain, the same domain we use in our first experiment: Grieser et al. [7] proposed an ontological similarity measure in cultural heritage collections, such as museums and art galleries. They combine the Wikipedia category hierarchy with lexical similarity measures to estimate the perceived relatedness of exhibits by museum visitors. Ensan et al. [8] proposed a semantic retrieval model to form a graph representation of documents and queries using semantic entity linking systems.

We used the *Semantic Connectivity Score* (SCS) [4] in our first experiment on comparing different museums. As detailed in Section 5.1.1, Graph-based and Wikipedia structure-based approaches do not suit well for comparing different museums described on Wikipedia.

We used the *Wikipedia Link-based Measure* (WLM) in the first experiment, as a baseline, described in Section 5 since it is a well-known metric to compare different Linked Data entities.

**Comparing Linked Data Entities with ontology and Wikipedia structure-based**

Another group of approaches focuses on machine learning techniques and Natural Language Processing (NLP) to compare entities.

Leme et al [9] proposed a technique based on probabilistic classifiers to the dataset recommendation problem. They ranked the most relevant datasets to recommend based on the probability that links between datasets can be found. Passant et al [10] measured semantic distances on Linked Data in order to provide a new kind of self-explanatory recommendations, combining Linked Data and Artificial Intelligence principles. They considered only the links that can exist between Linked Data resources, using both direct and indirect links to provide Linked Data resource recommendations. Morales et al. [11] proposed *MateTee*, a semantic similarity measure that combines the gradient descent optimization method with semantics encoded in ontologies to compute the relatedness among entities in Large Knowledge Graphs (KGs). Taieb et al. [12] proposed a method using the taxonomy of features extracted from ontologies to compute semantic similarity. Gabrilovich et al. [13] proposed a method that uses machine learning to compute the relatedness of texts fragments, called Explicit Semantic Analysis (ESA).

In this work, we do not use machine learning in the similarity computation. We consider in the similarity computation, the Linked Data entities to be compared, and their features, also Linked Data entities. Besides that, we do not use natural language processing (NLP) to compare text, since we consider the entities as the objects to be compared.

**Ranking Linked Data entities**

Several approaches for ranking Linked Data entities exploit the semantic aspects of Linked Data datasets.

Hogan et al. proposed ReConRank [14], a ranking method that adapts the well-known PageRank HITS algorithms to Semantic Web data. Roa-Valverde et al. [15] formalized the problem of ranking information in the Web of Data, unified the core concepts that characterize ranking algorithms, and compared different approaches to rank Linked Data. Mirizzi et al. [16] proposed the system *DBPediaRanker*, that explores the DBpedia graph and queries external sources, such as search engine results and social tagging systems, to compute a similarity value between the nodes.

Mirizzi et al [16] is the most relevant work to our approach. However, the strategies used to compare Linked Data Entities are slightly different: they use search engine results and social tagging system to compute the entities similarity, instead, we compare two Linked Data entities on comparing their (ranked) relevant features.

Our work fits in feature-based methods, described by Hliaoutakis et al. [17], that measure the similarity between two terms as a function of their properties. In this case, common features tend to increase the similarity between two concepts, or entities, and non-common features tend to diminish their similarity.

## 3   Preliminaries

As anticipated in the introduction, the similarity between two entities is estimated by comparing their ranked lists of features, using rank correlation metrics to generate the similarity score.

This section first reviews two measures to compare Linked Data entities: the *Semantic Connectivity Score* (SCS) [3], used in the first experiment as an attempt to compare museums, described in Section 5, and the *Wikipedia Link-based Measure* (WLM), used as baseline in the same experiment, also described in Section 5. Then, this section reviews two rank correlation metrics, the *Average Overlap* (AO) and the *Rank-biased Overlap* (RBO). The RBO metric is used in the three experiments as a rank correlation metric to compare the entities features. Finally, it reviews an information retrieval metric to measure the quality of ranks, the *Normalized Discounted Cumulative Gain* (NDCG), used in the first experiment to evaluate the results, and finally, it reviews relevant concepts related to the clustering task.

*Wikipedia Link-based Measure* (WLM) and *Semantic Connectivity Score (SCS)*

In the literature, there are several measures to compare Linked Data entities [3][18]. In particular, Milne and Witten proposed, in 2003, the *Wikipedia Link-based Measure* (WLM) [2]. Formally, the relatedness between two Wikipedia articles of interest *a* and *b* is defined as follows:

$$sr(a,b) = \frac{\log(\max(|A|,|B|)) - \log(|A \cap B|)}{\log(|W|) - \log(\min(|A|,|B|))}$$

where $A$ and $B$ are the sets of all articles that link to $a$ and $b$, respectively, and $W$ is the entire Wikipedia.

Another measure is the *Semantic Connectivity Score* (SCS), proposed by Nunes et al. [3]. The SCS between a pair of entities $a$ and $b$ is based on the Katz score [4] and is defined as follows:

$$SCS(a,b) = \sum_{l=1}^{\tau} \beta^l \cdot |paths_{(a,b)}^{<l>}|$$

where $|paths_{(a,b)}^{<l>}|$ is the number of paths between entities $a$ and $b$ of length $l$, $\tau$ is the maximum path length considered and $\beta$ is a positive damping factor, ranging from 0 to 1, responsible for exponentially penalizing longer paths.

*Average Overlap* (AO) and *Rank-biased Overlap* (RBO)

Correlating ranked lists is a common problem in several areas, such as graph analysis and information retrieval. Webber et al. [19] categorize them according to two main characteristics: the *conjointness* (two conjoint lists consist of the same items) and the *weightedness* (a list is *weighted* when the items have different relevance and a list is *top-weighted* when the top of the list is more important than the tail).

For conjoint lists, widely used rank correlation coefficients are Kendall's and Spearman's [20]. For non-conjoint lists with items of different weights (ranks), there are similarity measures that can be used, such as *Jaccard* [21], *Cosine Similarity*, and *Average Overlap* [22][23], and the *Rank-biased Overlap* (RBO). When dealing with non-conjoint lists, it is common to start from the set intersection, considering the size of the intersection or the overlap between the two rankings. The *Average Overlap* (AP) is based on the set intersection, but considers the overlap at increasing depths when comparing two rankings, and is defined as follows [19]:

$$AO(S,T,k) = \frac{1}{k} \sum_{d=1}^{k} A_d$$

where $S$ and $T$ are two possibly infinite lists, $k$ is the evaluation depth, and $A_d$ is their agreement at depth $d$, defined as

$$A_d = \frac{1}{d} |S_d \cap T_d|$$

where $S_d$ (or $T_d$) is the prefix of $S$ (or $T$) up to depth $d$. For each d ∈ {1,…,k}, it calculates the overlap at $d$, and then averages those overlaps to derive the similarity measure.

Table 1 gives a sample calculation of AO. Up to depth 2, the AO score is 0 since the common items in $S$ and $T$ first appear at the depth 3. If we consider, for instance, depth 7, AO gives 0.312 as the similarity score between $S$ and $T$. Although, the higher AO score is at depth 6, 0.317.

**Table 1:** *Average Overlap (AO) of two lists* [19]

| $d$ | $S_d$ | $T_d$ | $A_d$ | $AO(S,T,d)$ |
|---|---|---|---|---|
| 1 | <a> | <z> | 0.000 | 0.000 |
| 2 | <ab> | <zc> | 0.000 | 0.000 |
| 3 | <abc> | <zca> | 0.667 | 0.222 |
| 4 | <abcd> | <zcav> | 0.500 | 0.292 |
| 5 | <abcde> | <zcavw> | 0.400 | 0.313 |

| d | $S_d$ | $T_d$ | $A_d$ | AO(S,T,d) |
|---|---|---|---|---|
| 6 | <abcdef> | <zcavwx> | 0.333 | 0.317 |
| 7 | <abcdefg> | <zcavwxy> | 0.286 | 0.312 |
| k | <abcdefg...> | <zcavwxy...> | ... | ... |

Webber et al. [19] proposed to extend this idea to incomplete ranks (i.e. they do not cover all elements in the domain) by adding a parameter that determines the importance of the weighting of the top ranks. They defined the *Rank-biased overlap* (RBO) as follows:

$$\text{RBO}(S,T,p) = (1-p)\sum_{d=1}^{\infty} p^{d-1} \cdot A_d$$

The RBO measure handles non-conjoint lists and weights higher ranks more heavily than lower ranks (their top-k item is more relevant than the top-k+1, and so on). In addition, RBO has a parameter *p*, which ranges from 0 to 1, and determines the strength of the weighting of the top ranks, i.e., the smaller *p*, the more top-weighted the metric is. If *p* = 0, only the top-ranked item is considered, and the score is either 0 or 1. Rank-biased overlap ranges from 0 to 1, where 0 means disjoint, and 1 means identical.

Although the ranked lists we use in the experiments are not incomplete, we use RBO as the main rank correlation metric for non-conjoint lists, since it allows imposing a stronger penalty on differences at the top of the ranking than on differences further down, the lists.

In this paper, we compare RBO with different strategies, depending on the scenario. In the first experiment that compares museums, we compare RBO results with the semantic relatedness measure WLM (*Wikipedia Link-based Measure*) [2]. Although this article focuses on the semantic *similarity* between Linked Data entities, we used the semantic *relatedness* measure WLM as a baseline, since it is a well-known metric to compare different Linked Data entities.

We used WLM as a baseline neither for the second experiment (comparing datasets) nor for the third experiment (comparing conferences), because WLM uses the Wikipedia structure (or its RDF version) to compute the similarity, and yet these instances do not have an entry in Wikipedia (and hence in DBpedia). Instead, in the second and third experiments, we compared RBO both with the Jaccard and cosine distances, other widely used similarity metrics. Section 4 describes the experiments and their results.

*Normalized Discounted Cumulative Gain* (NDCG)

Traditionally used in search engine results to evaluate ranks, NDCG (*Normalized Discounted Cumulative Gain*) [24] emphasizes retrieving highly relevant documents.

Intuitively, the idea behind NDCG is that a recommender system returns some items and we would like to evaluate how good the list is. Each item (or document) has a relevance score (or *gain*) that are added up (*cumulative* gain). Since we prefer to find the most relevant items at the top of the list, before summing the gains, they are divided by a growing number (usually a logarithm of the item position) – that is, *discounting*. Since DCG are not directly comparable between users, we *normalize* it.

Let $g_1$, $g_2$, ..., $g_Z$ be the gain values associated with the *Z* documents retrieved by a system in response to a query. Let $DCG_I$ denotes the DCG value for an ideal ranked list for the query. NDCG is defined as follows [24]:

$$nDCG = \frac{DCG}{DCG_I} \text{ where } DCG = \sum_{i=1}^{z} g_i / \log(i + 1)$$

NDCG returns a non-negative score ranging from 0 to 1 for the first *k* items on the list (NDCG@*k*), since a recommender system is interested in few items, to be considered as relevant and be shown to the users.

In the first experiment, we use NDCG to evaluate the results, since we compare lists of similar museums generated by our approach with the lists generated by the ground truth. However, in the second and in the third experiment, we chose clustering techniques to evaluate the results. Clustering algorithms group a set of objects in a way that objects in the same group (or cluster) are more similar to each other than to objects in other clusters. We assume, therefore, that the most similar datasets (or conferences) should belong to the same cluster.

*Clustering performance evaluation*

There is a wide variety of clustering algorithms (for instance, hierarchical agglomerative, centroid-based, among others). Besides the similarity measure used to compare the objects (that can be for instance *Jaccard*, *Cosine* or *RBO*), such algorithms also depend on a method for grouping the objects, called *linkage criteria*, listed as follows:

- *Single Linkage* minimizes the minimum distance criterion between items in pairs of clusters (see Figure 1(a)).
- *Complete Linkage* minimizes the maximum distance between items in clusters (see Figure 1(b)).
- *Average Linkage* minimizes the maximum distance between items in clusters but considering the average of the distance between the items (see Figure 1(c)).
- *Ward Linkage* minimizes the sum of squared differences between items.

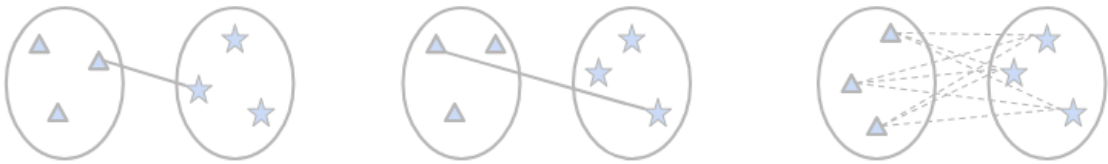

**Figure 1.** (a) Single Linkage; (b) Complete Linkage; (c) Average Group

Another important task is the evaluation of the clustering algorithm. The *Rank index* (RI) computes the similarity between two clusters by considering all pairs of items and counting pairs that are assigned to the same or different clusters in the predicted and in the true cluster and lies in the range of [0,1]. In practice, however, RI often ranges from 0.5 to 1. Also, its baseline value can be high and does not take a constant value. For these reasons, RI has been mostly used in its adjusted form [25], known as the *Adjusted Rand Index* [26], which ranges from -1 and 1.

In the experiments to evaluate the quality of the generated clusters, when comparing them with the clusters of the ground truth, we adopted the Adjusted Rand Index (ARI).

## 4 Proposed method to compute entity semantic similarity

In this section, we first present two basic definitions: the *ranked features* and the *entity similarity*. Then, we present a fragment of a framework called SELEcTor, both first introduced by Ruback et al. [1].

### 4.1 Basic definitions

The problem of discovering similar entities on Linked Data by ranking and comparing their features depends on two basic definitions.

DEFINITION 1 (RANKED FEATURES). The *ranked features* of a Linked Data entity $e$ is a list $F = ((f_1,s_1),…,(f_n,s_n))$, where $f_j$ is a feature of $e$ and $s_j$ is the *relevance score* of $f_j$, for $j \in [1, n]$.

Intuitively, feature $f_j$ is more relevant to describe $e$ than feature $f_{j+1}$, for $1 \leq j \leq n$.

DEFINITION 2 (ENTITY SIMILARITY). Given two Linked Data entities $e_i$ and $e_j$ and their ranked features $F_i$ and $F_j$, the *similarity* between $e_i$ and $e_j$ is the distance between $F_i$ and $F_j$, according to a rank correlation metric $m$.

Note that entities $e_i$ and $e_j$ may have a different number of relevant features, i.e., their lists $F_i$ and $F_j$ may have different sizes. Furthermore, $F_i$ and $F_j$ may or may not have features in common.

### 4.2 The SELEcTor framework

The SELEcTor framework (illustrated in Figure 2) [1] takes as input Linked Data entities, extracts their relevant features from datasets found in the Linked Open Data cloud, and then compares the ranked features according to some rank correlation metric to generate the entities similarity score. The first phase is the *ranked features extractor*, which communicates with Linked Open datasets to extract the ranked lists of relevant features that describe the entities. The second phase, the *entity similarity processor*, takes these lists as input and compares them using a list correlation metric to generate as output the entity similarity score. In the following sections we detail the two phases.

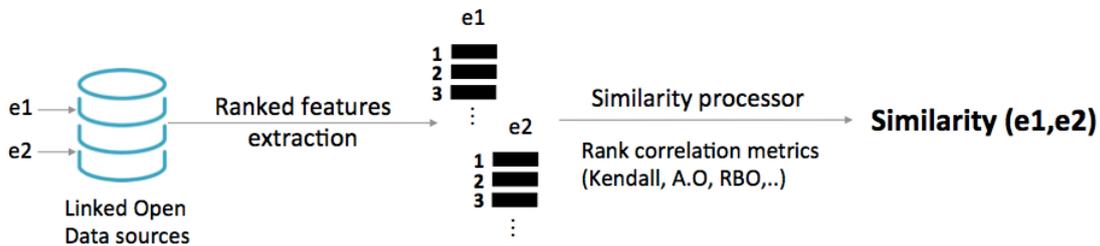

Figure 2. A fragment of the SELEcTor framework

#### 4.2.1 Extracting ranked features

The first phase of the SELEcTOR framework extracts the ranked features generating a ranked lists of features to describe the entities to be compared. As it can be seen in Figure 2, this module receives as input two Linked Data entities and outputs their respective ranked features. It is important to notice that feature identification is part of an analysis process that can be aided by a domain expert.

Taking an entity as input, the module can generate ranked lists using two different strategies: (i) navigating on the Linked Data graphs through the nodes connected to the entity to extract the features or (ii) accessing the Linked Data graphs through their respective SPARQL endpoints. The first strategy is referred to as the *graph-exploration* approach and the latter is referred to as *query-based* approach by Ruback et al. [1].

The *query-based* approach performs a pre-defined SPARQL query over one or more Linked dataset endpoint to generate the ranked features. The relevance criterion is applied using some group function that aggregates the features. We note that, depending on the aggregation function chosen, a tie may happen between two or more features, in the sense that they have the same number of feature values. In these cases, the SPARQL query can be re-formulated to untie the elements according to some other criteria. In Section 5.1.1, we present an example of a SPARQL query for the museum's domain.

When applying the *graph-exploration* approach, the module navigates through the RDF graph and then calculates the importance of each node to describe a certain entity. There are several approaches in the literature to measure relationships within a graph, referred as centrality measures, such as the Katz score [4] or the SCS score [3]. They can be profitably applied in this context to generate a ranked list of features (nodes) that represents a certain entity. In Section 5.1.1, we present an example of using the SCS score in the museum's domain.

### 4.2.2 Computing entity similarity

The *similarity processor* is the second phase shown in Figure 2, and takes as input the lists of ranked features and compares them to measure how similar they are, using a rank correlation metric. The module outputs the similarity score for the pair of entities.

The module can choose a measure to compute the similarity between the entities, according to the nature of the ranked features. When the lists have the same items (i.e., if they are conjoint), the module may choose Kendall, Spearman's p, among other metrics [27]. Otherwise, the module may choose AO (Average Overlap) [23] or its top-weighted parameterized extension, RBO (see Section 3).

The module outputs a similarity score that measures how similar the entities are. When the entities do not have any feature in common, the correlation coefficient is 0, and when they have the same features exactly in the same order (and the same weights for weighted ranks), the output is 1.

In the next section, we instantiate the SELEcTor framework in three different domains to evaluate the approach.

## 5 Evaluating the proposed approach to estimate entity similarity

To assess the proposed approach to estimate entity similarity, we evaluate SELEcTOR in three different domains: (i) museums descriptions found in DBpedia; (ii) datasets published as Linked Data and described in a Linked Open Data catalogue; and (iii) computer science conferences available in the Linked Data version of DBLP. For each domain, we used a ground truth and a baseline approach to compare the proposed entity similarity measure with.

In the museum's domain, since there is no specific ground truth, we constructed a ground truth from a curated Web site about art history, which is independent of DBpedia;

for each museum, the ground truth associates a ranked list of similar museums. Then, for each museum, we generated a ranked list of similar museums based on their ranked features. Finally, we compared such lists with the ground truth lists using NDCG (see Section 3).

In the LOD datasets domain, we considered as ground truth the manual classification of the LOD datasets into communities, expressed in the LOD cloud[2] diagram, and reinterpreted as *ground truth dataset clusters*. For each dataset, we generated a ranked list of features using the DBPedia top-level categories. Finally, we clustered the datasets, using the proposed similarity measure based on the ranked feature lists, and compared the clusters thus obtained with the ground truth dataset clusters. Assuming that the most similar datasets should belong to the same cluster, the experiment uses a clustering performance evaluation metric.

In the computer science conferences domain, we considered as ground truth the conferences defined in Wikipedia, already organized in groups (or clusters) of similar conferences. For each conference, we generated a ranked list of features using the papers keywords. Again, we clustered the conferences, using the proposed similarity measure based on the ranked feature lists, and compared the clusters thus obtained with the ground truth clusters, as for the second experiment.

The next sections describe in detail the set of features adopted in each domain, the method to extract them, and the results of the experiments.

### 5.1.1 Comparing museums in DBpedia

*Museum entities in DBpedia*

DBpedia contains entities that represent museums, which are instances of the `dbo:Museum` class. An example is the triple `<dbr:Louvre, rdf:type, dbo:Museum>` (dbr is the prefix of `http://dbpedia.org/r.esource/)` stating that the Louvre is an entity of type `dbo:Museum`.

Museum instances can be linked to other entities through the `dct:subject` property, often used to represent the topic of the entity. Some of these entities are hierarchically related to each other through the `skos:broader` property and may have a direct link to the `dbc:Museum_by_type` class. We call *categories* all the entities linked to `dbc:Museum_by_type` directly or indirectly through the `skos:broader` property.

Figure 3 shows the categories associated with the Louvre. A category directly related to Louvre is `dbc:Museums_of_Ancient_Greece`. The indirectly related categories are `dbc:History_museum` and `dbc:Civilization_museums`.

---

[2] http://lod-cloud.net/

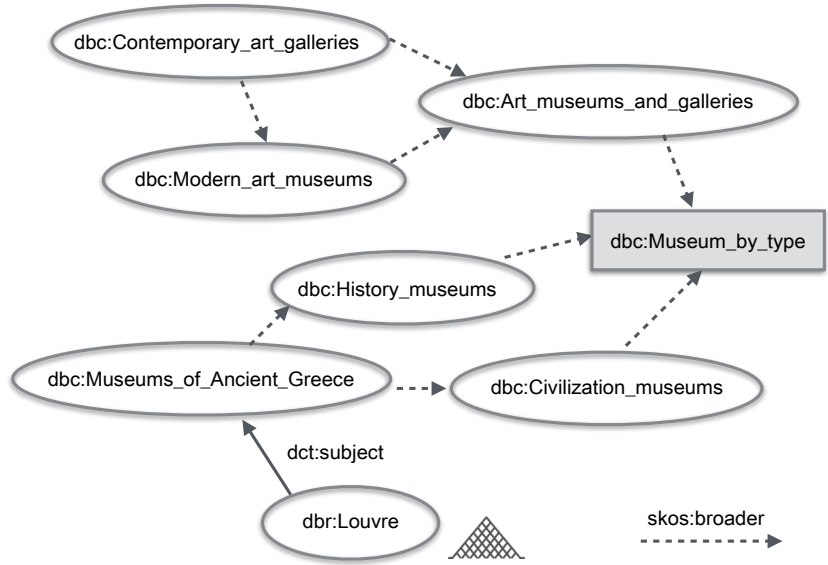

**Figure 3**. DBpedia concepts describing museum categories.

Figure 4 illustrates other museum properties. The `dbo:museum` property links a museum to its artworks, which are instances of `dbo:Artwork`. In turn, each artwork is linked to its creator/artist through the `dbo:author` property. Finally, the artists are related to one or more art movements by the `dbo:movement` property. The RDF graph shown in Figure 3 represents that the *J Paul Getty Museum* has as artwork the *Irises* painting by *Vincent van Gogh*, a post-impressionist (art movement) artist.

In the experiments, we explored both DBpedia graphs, shown in Figure 3 and Figure 4, as follows.

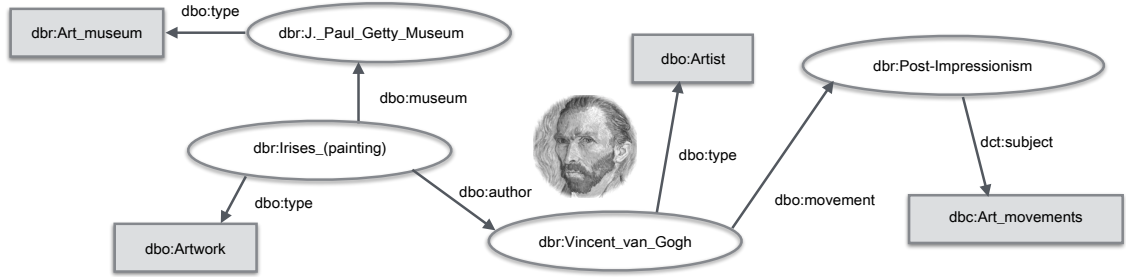

**Figure 4**. DBpedia links describing J. Paul Getty museum features.

*Extracting and ranking features*

According to the first step of the SELECTOR framework – "extraction of ranked features", we explored the DBpedia entities and properties shown in Figure 3. We navigated through the graph first using the *graph-exploration approach* (see Section 4.2.1), from the root entity (the museum) reaching the entities that represent their categories.

We consider such museum categories (the entities having direct or indirect links to the `dbc:Museum_by_type` class) as the features to be ranked. We used the depth-first approach with depth distance 4, as adopted by Fang et al. [28], which means that we considered entities from the root until reaching all its 4-hop neighbors.

To measure the relevance of each feature with respect to the museum, we computed the distance from the museum to all the features (the categories) using the SCS score (see Section 3). This strategy We then ordered the features according to the score, generating the ranked features list.

Table 2 shows the Louvre ranked features, using the graph-exploration approach, ordered in descending order by their SCS score. The feature `dbc:Museums_of_Ancient_Near_East`, at the first position, and the feature `dbc:Museums_of_Ancient_Greece`, at the second position, are the most relevant features to describe the Louvre (both with SCS score 0.5), while the less relevant is `dbc:Civilization_museums`, with SCS score 0.25.

Table 2. Louvre features using museums hierarchy of categories

| Louvre categories | SCS score |
|---|---|
| dbc:Museums_of_Ancient_Near_East | 0.5 |
| dbc:Museums_of_Ancient_Greece | 0.5 |
| dbc:History_museums | 0.48 |
| dbc:Civilization_museums | 0.25 |

We found that the ranked features based on the museums hierarchy of categories (shown in the graph in Figure 3) did not represent the museums suitably, for two reasons: (i) some of the most famous museums have few DBpedia categories that represent them, such as the Louvre museum; and (ii) the categories found are often very generic and thus do not represent the museums appropriately (such as `dbc:Society_museums`, `dbc:Art_museums_and_Galleries` and `dbc:Civilization _museums`).

We then proceeded to the *query-based* approach (see Section 4.2.1), to explore the DBpedia graph paths shown in Figure 4. Instead of exploring the museum categories, we focused on the art movements of the artworks that can be reached using the artists, shown in Figure 4. The *ranked features extractor* module performed the following SPARQL query, which aggregates the art movements and orders them by the artwork frequency. The `?museum` parameter represents the input entity.

```
SELECT ?artMovement
WHERE {
  ?artWork <dbo:museum> ?museum.
  ?artWork <dbo:author> ?artist.
  ?artist <dbo:movement> ?artMovement.
}
GROUP BY ?art_movement
ORDER BY DESC(count(?artWork))
```

Table 3 shows the SPARQL query results for the Getty Museum. Note that, since DBpedia is constantly updated, the results vary with time. In some cases, the SPARQL query returns items with the same feature value, i.e., two or more art movements with the same artwork frequency. In this case, one may choose another criterion to untie these features, such as the *out* or the *in* degree of the feature, which respectively represents the number of RDF out-links leaving from the entity and the number of in-links pointing to the entity.

Table 3. Ranked list of features for the Getty Museum, using the artwork frequency.

| Getty Museum ranked list of features |
|---|
| dbr:Symbolism_(arts) |
| dbr:Baroque |
| dbr:Expressionism |
| dbr:Romanticism |
| dbr:High_Renaissance |
| dbr:Dutch_Golden_Age_painting |
| dbr:Academic_art |
| dbr:Post-Impressionism |
| dbr:Mannerism |

The feature that best describes the Getty Museum is the *Symbolism* art movement, while the less relevant feature is the *Mannerism* art movement. This means that the museum has more artworks of the *Symbolism* art movement than of the Mannerism art movement.

Comparing these two strategies to extract features for the museum's domain, we found that the latter strategy (referred to as *query-based* approach) can extract better features, both in quantity (the ranked lists have more items) and in quality (the art movements are more domain-specific than the museum categories available in DBpedia).

*Computing entity similarity*

As explained before, the *entity similarity processor* module takes as input the ranked lists of features, representing the entities to compare, and outputs a similarity score. Table 4 shows the ranked features that describe the Getty and the Louvre museums (for a matter of space, some Louvre features have been omitted). They have been generated by performing the SPARQL query previously presented.

Table 4. Comparing the ranked features.

| J Paul Getty ranked features | Louvre ranked features |
|---|---|
| dbr:Symbolism_(arts) | dbr:Romanticism |
| dbr:Baroque | dbr:High_Renaissance |
| dbr:Expressionism | dbr:Neoclassicism |
| dbr:Romanticism | dbr:Baroque |
| dbr:High_Renaissance | dbr:Italian_Renaissance |
| dbr:Dutch_Golden_Age_painting | dbr:Dutch_Golden_Age_painting |
| dbr:Academic_art | dbr:The_Renaissance |
| dbr:Post-Impressionism | dbr:Classicism |
| dbr:Mannerism | dbr:Realism_(arts) |
|  | dbr:Flemish_Baroque_painting |
|  | dbr:Early_Netherlandish_painting |
|  | dbr:Caravaggisti |

To compare the two ranked lists of features, the *entity similarity processor* module applies the RBO measure (see Section 3), which handles non-conjoint lists (the museums are not necessarily described by the same art movements) and weights high ranks more heavily than low ranks (their top-k art movement is more relevant then the top-k+1, and so on).

When computing the similarity between the Getty and the Louvre museums, with the top-weighted parameter $p = 0.95$, the RBO score is 0.437. In fact, from a total of 4 common features (the art movements), the first art movement that occurs in both the Getty and in the Louvre lists (`dbr:Baroque`) appears in the 4$^{nd}$ position in the Louvre list, and the last art movement to match (`dbr:Dutch_Golden_Age_Paiting`) appears in the 6th position in the Louvre list.

When comparing the Getty Museum with the Museum of Modern Art, in New York, the RBO similarity score, with p = 0.95, is 0.117. In fact, they only have 2 common art movements, the first art movement matches in the 8th position (`dbr:Post-Impressionism`) and the last art movement matches in the 14th position (`dbr:Expressionism`). Considering our museum's dataset shown in Table 5, the Art Institute of Chicago is the museum least similar to the Getty Museum.

We computed the similarity between all museums of our dataset (presented in Table 5) using RBO. Then, we generated, for each museum, the list of the most similar museums. Table 5 shows the most similar museums to the Getty Museum, with $p = 0.95$ and $p = 0.98$.

Table 5. The museums most similar to J. Paul Getty.

| Getty similars | RBO score p = 0.95 | RBO score p = 0.98 |
|---|---|---|
| dbr:Metropolitan_Museum_of_Art | 0.437 | 0.491 |
| dbr:Louvre | 0.404 | 0.429 |
| dbr:Kunsthistorisches_Museum | 0.385 | 0.419 |
| dbr:Museum_of_Fine_Arts,_Boston | 0.360 | 0.381 |
| dbr:Vatican_Museums | 0.351 | 0.380 |
| dbr:Uffizi | 0.261 | 0.302 |
| dbr:National_Gallery_of_Art | 0.247 | 0.281 |
| dbr:Musée_d'Orsay | 0.161 | 0.195 |
| dbr:Philadelphia_Museum_of_Art | 0.161 | 0.195 |
| dbr:Museum_of_Modern_Art | 0.117 | 0.139 |
| dbr:Art_Institute_of_Chicago | 0.103 | 0.130 |

*A ground truth for the museum's domain*

Since there is no specific ground truth containing museums similarity data to validate our approach, we built a ground truth using a well-known Web site about art history, *Smart History*[3], which is a non-profit organization that makes art history learning content freely available, and provides several articles discussing the most important masterpieces, ranging from ancient to contemporary art. We chose *Smart History* as the ground truth because it is a rich source of museums data entirely authored by human domain experts, its creation process is totally independent of DBpedia (or similar) data, and it is not affected by popularity bias.

Each *Smart History* article (very often an article is about an artwork) is categorized according to a hierarchical taxonomy, which includes time periods, art movements, and other relevant facets. For each artwork, the hosting museum is also mentioned.

Given the *Smart History* data, we defined the similarity between two museums based on the categories found in the articles mentioning the museums. To avoid sparsity, we limited to the top-2 levels of the category hierarchy. Museum similarity was then computed as the cosine similarity of the museum's categories. Cosine similarity was adopted as it allows to properly weight the richness of a given museum in a specific category, but it also avoids boosting large museums with many artworks.

---
[3] http://smarthistory.org

*Evaluating the results*

First, we pre-filtered 32 museums in DBpedia with at least 8 artworks and 5 art movements to avoid poorly described museums. Then, we filtered the museums that also have categories in the ground truth, *Smart History*, resulting in the 12 richest museums in both datasets, shown in Table 6. From the Louvre museum, for instance, one can reach 17 art movements in DBpedia and one can find 37 art movements in the Smart History Web site.

Table 6. Chosen museums for the experiment.

| Museum | #DBpedia art movements | #SmartHistory art movements |
|---|---|---|
| dbr:Metropolitan_Museum_of_Art | 28 | 84 |
| dbr:Louvre | 17 | 37 |
| dbr:Museum_of_Modern_Art | 36 | 17 |
| dbr:National_Gallery_of_Art | 29 | 17 |
| dbr:J._Paul_Getty_Museum | 9 | 14 |
| dbr:Uffizi | 11 | 12 |
| dbr:Museum_of_Fine_Arts,_Boston | 7 | 16 |
| dbr:Musée_d'Orsay | 11 | 10 |
| dbr:Art_Institute_of_Chicago | 26 | 9 |
| dbr:Philadelphia_Museum_of_Art | 15 | 13 |
| dbr:Kunsthistorisches_Museum | 9 | 11 |
| dbr:Vatican_Museums | 16 | 13 |

We chose as a baseline the semantic relatedness measure proposed by Milne and Witten, WLM (see Section 3). Even though WLM is intended to be a generic approach, we chose it as a baseline, since it measures the semantic relatedness of two Linked Data entities. To compute the WLM similarity, we used Dexter, an Open Source Framework for Entity Linking [18].

Our strategy to evaluate the results is based on a comparison of the lists of similar museums (such as in Table 5) generated by the three different approaches: (i) our approach; (ii) the WLN measure, which represents our baseline; and (iii) the Smart History data, from which we built the ground truth.

As an example, Table 7 shows the lists of museums which are similar to the Getty Museum, as generated by our approach, and using Smart History. As shown in Table 7, according to our approach, the *MET* (Metropolitan Museum of Art, in New York) is the museum most similar to the *Getty Museum*, the second most similar is the Louvre Museum, and the third one is the *Kunsthistorisches* Museum, an art museum in Vienna, and so on.

We also compared the lists of similar museums generated by WLM (our baseline) and by *Smart History*. As an example, Table 8 shows the lists of museums which are similar to the *Getty Museum* as generated by WLM and by *Smart History*. As shown in Table 8 the most similar museum to the Getty Museum is the *National Gallery of Art*, an art museum in Washington D.C. The second most similar museum is the *Musée d'Orsay*, in Paris, and the third most similar is the *Philadelphia Museum of Art*, and so on.

Table 7. Comparing our approach with the ground truth.

| Museums similar to the Getty (SELEcTor) | Museums similar to the Getty (using SmartHistory data) |
|---|---|
| dbr:Metropolitan_Museum_of_Art | dbr:Metropolitan_Museum_of_Art |
| dbr:Louvre | dbr:Vatican_Museums |
| dbr:Kunsthistorisches_Museum | dbr:Louvre |
| dbr:Museum_of_Fine_Arts,Boston | dbr:National_Gallery_of_Art |
| dbr:Vatican_Museums | dbr:Art_Institute_of_Chicago |
| dbr:Uffizi | dbr:Museum_of_Fine_Arts,Boston |
| dbr:National_Gallery_of_Art | dbr:Musée_d'Orsay |
| dbr:Musée_d'Orsay | dbr:Philadelphia_Museum_of_Art |
| dbr:Philadelphia_Museum_of_Art | dbr:Kunsthistorisches_Museum |
| dbr:Museum_of_Modern_Art | dbr:Uffizi |
| dbr:Art_Institute_of_Chicago | dbr:Museum_of_Modern_Art |

Table 8. Comparing WLM with the ground truth.

| Museums similar to the Getty (WLM approach) | Museums similar to the Getty (using SmartHistory data) |
|---|---|
| dbr:National_Gallery_of_Art | dbr:Metropolitan_Museum_of_Art |
| dbr:Musée_d'Orsay | dbr:Vatican_Museums |
| dbr:Philadelphia_Museum_of_Art | dbr:Louvre |
| dbr:Museum_of_Fine_Arts,Boston | dbr:National_Gallery_of_Art |
| dbr:Kunsthistorisches_Museum | dbr:Art_Institute_of_Chicago |
| dbr:Art_Institute_of_Chicago | dbr:Museum_of_Fine_Arts,Boston |
| dbr:Metropolitan_Museum_of_Art | dbr:Musée_d'Orsay |
| dbr:Uffizi | dbr:Philadelphia_Museum_of_Art |
| dbr:Museum_of_Modern_Art | dbr:Kunsthistorisches_Museum |
| dbr:Vatican_Museums | dbr:Uffizi |
| dbr:Louvre | dbr:Museum_of_Modern_Art |

Analyzing the results, we note that the geographic proximity between two museums influences the similarity score between them, according to WLM, as expected. This is because this measure considers all links found in their respective Wikipedia articles, including some geographic-related links. An example is the link `<dbc:Modern_art_museums_in_the_United_States>`, connected to the museum through the property `<dct:subject>`, that can be found both in the *Metropolitan Museum* page and in the National Gallery of Art page (`dbc` is a prefix for `http://dbpedia.org/page/Category` and `dct` is a prefix for `http://purl.org/dc/ terms/`). This explains why, according to WLM, some museums – for instance the *Louvre* and the *Vatican* – are in the last positions in the similarity list in Table 8, while they appear in the first half of the list, in SELEcTor list of similar museums in Table 7.

Analogously, WLM considered as similar to the Getty Museum some museums that SELEcTor does not – for instance, the *Philadelphia Museum of Art*, which is also located in the United States.

Lastly, we calculated the accuracy of the SELEcTor lists and the WLN lists by comparing both with the ground truth. We compared the lists using NDCG [29], a well-known metric used in Information Retrieval to measure ranking quality (see section 3). This measure accumulates the gain from the top of the list to the bottom, penalizing lower ranks. It can be parameterized to consider the top k elements of the lists.

Figure 5 shows the results considering only the *Getty Museum*, with *k* from 3 to 8. Considering the top 3 items, the SELEcTor accuracy score, using RBO, is 0.886, while the WLM score is 0.697. Considering the top 4 items, the SELEcTor score decreases to

0.876, but it is still higher than the WLM score, 0.714. The highest accuracy is 0.924, achieved when *k* = 8.

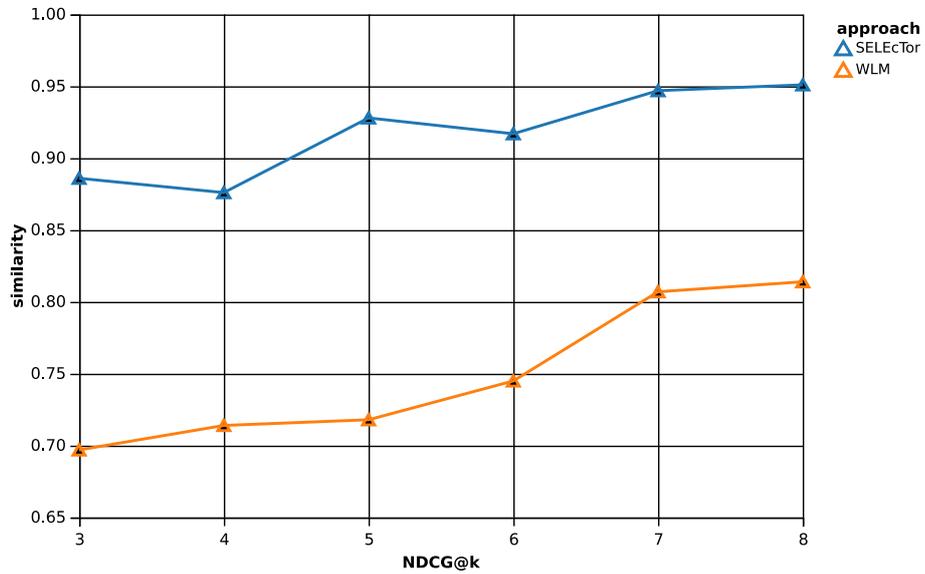

**Figure 5.** NDCG results for Getty museum.

Finally, we compared the similarity lists of all museums using the same idea, with *k* again ranging from 3 to 8. Figure 6 shows the results. Considering all museums, SELEcTor performs significantly better than WLM, for any *k*, with the highest accuracy being 0.924, when *k* = 6.

Indeed, since SELEcTor filters the entities that better describe a museum, it can be considered more selective than WLM, in the sense that it focuses on the specific features that describe the museum, such as its art movements.

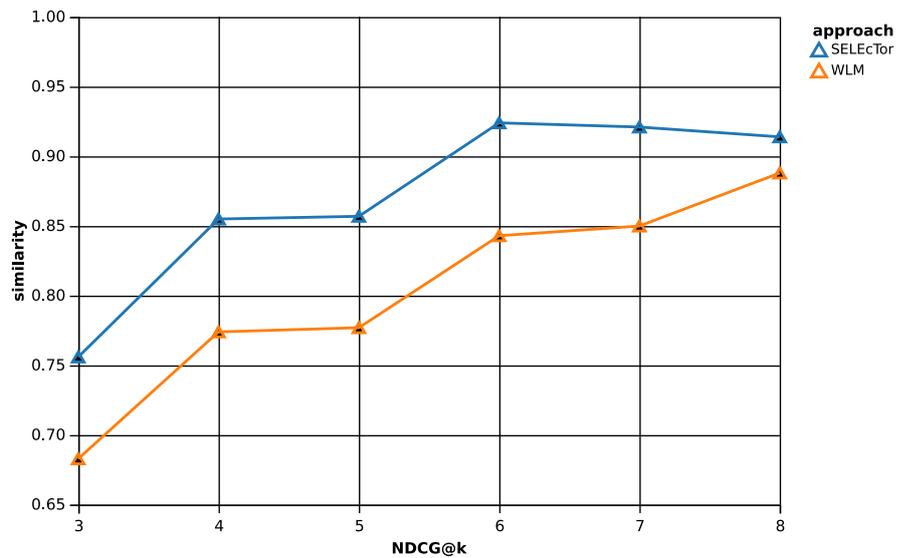

**Figure 6.** The average NDCG top k items.

### 5.1.2   Comparing LOD datasets

*Datasets in the Mannheim Linked Data Catalog*

The Mannheim Linked Data Catalog[4] is a popular catalog that contains information about datasets that are published as Linked Data on the Web and was generated by a crawling process of the LOD cloud, detailed by Kawase et al. [30].

Figure 97 illustrates the overview of the experiment. The upper part of the figure comprises the Mannheim Linked Data Catalog extraction process. We extracted 390 datasets from the Mannheim catalog, filtering out only the datasets that have SPARQL endpoints available.

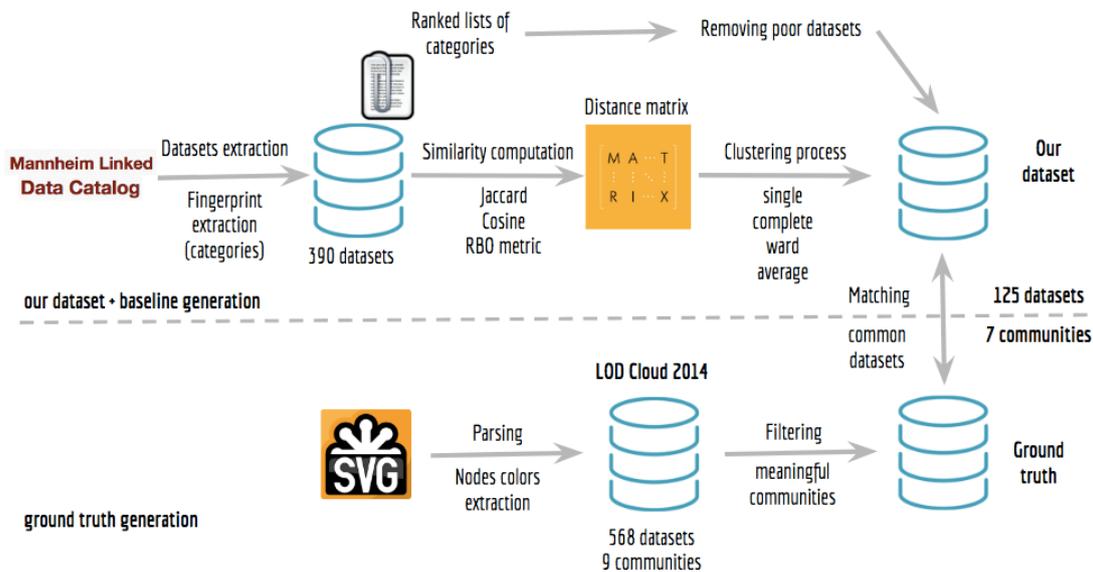

**Figure 7:** Datasets selection

*Extracting and ranking features*

There are several ways to extract relevant features from datasets. In this experiment, we adopted a profiling technique described by Caraballo et al. [31], which basically generates *profiles* or *fingerprints* for textual resources, extracted from the datasets (see Figure 7), detailed as follows.

1. Extract entities from a given textual resource.
2. Link the extracted entities to the English Wikipedia articles.
3. Extract the English Wikipedia categories for the articles.
4. Follow the path from each extracted category to its top-level category and compute a vector with scores for the top-level categories thus obtained.
5. Perform a linear aggregation in all dimensions of the vectors to generate the final profile, represented as a histogram for the 23 top-level categories of the English Wikipedia, shown in Table 9 (`dbc` is a prefix for `https://en.wikipedia.org/wiki/Category:Main_topic_classifications`).

---
[4] http://linkeddatacatalog.dws.informatik.uni-mannheim.de/

The Wikipedia top-level categories shown in Table 9 represent the features that describe the datasets. For each of the 390 datasets, we then aggregate the categories by counting the number of entities extracted from textual resources of each category.

**Table 9**. Wikipedia Top-level categories

| Wikipedia Top-level categories |
|---|
| dbc:Agriculture |
| dbc:Applied_sciences |
| dbc:Arts |
| dbc:Belief |
| dbc:Business |
| dbc:Chronology |
| dbc:Culture |
| dbc:Education |
| dbc:Enviroment |
| dbc:Geography |
| dbc:Health |
| dbc:History |
| dbc:Humanities |
| dbc:Language |
| dbc:Law |
| dbc:Life |
| dbc:Mathematics |
| dbc:Nature |
| dbc:People |
| dbc:Politics |
| dbc:Science |
| dbc:Society |
| dbc:Technology |

Each dataset is described by until 23 features, i.e., categories. The datasets were described by an average of 17.2 categories, with high variance and standard deviation, respectively 20.73 and 4.55. Figure 8 shows the frequency distribution of categories per dataset.

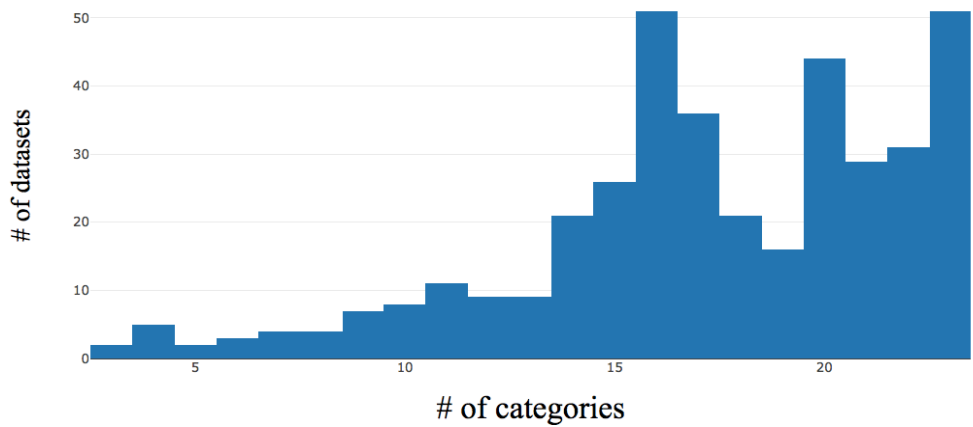

**Figure 8:** Frequency distribution of categories per dataset.

Finally, we avoided poorly described datasets, i.e., datasets described by a few categories. We considered a minimum of 15 categories (out of 23), which ruled out 10% of the datasets (the left part of the plot shown in Figure 8).

*Computing entity similarity*

We generate ranked lists using the frequency of each category found for the dataset. Table 10 shows the frequency of the categories for two of the datasets available at the catalog: the *eu-agencies-bodies* dataset[5], a dataset about agencies and decentralized bodies in the EU; and the *rkb-explorer-citeseer*[6] dataset, a semantic research repository with co-reference information from the research index CiteSeer[7]. Table 10 shows the top 5 categories of each dataset in boldface.

Table 10. Top-level categories frequency.

| Top-level category | *eu-agencies-bodies* dataset | *rkb-explorer-citeseer* dataset |
|---|---|---|
| dbc:Agriculture | 133 | 1 |
| dbc:Applied_science | 57 | 28 |
| dbc:Arts | 85 | 75 |
| dbc:Belief | 290 | 930 |
| dbc:Business | **687** | 468 |
| dbc:Chronology | 163 | **1377** |
| dbc:Culture | 427 | 773 |
| dbc:Education | 508 | **1458** |
| dbc:Enviroment | 411 | 91 |
| dbc:Geography | 240 | 14 |
| dbc:Health | 38 | 3 |
| dbc:History | 20 | 33 |
| dbc:Humanities | 460 | 861 |
| dbc:Language | 84 | 367 |
| dbc:Law | 15 | 2 |
| dbc:Life | 291 | 401 |
| dbc:Mathematics | 248 | **2142** |
| dbc:Nature | **1650** | 893 |
| dbc:People | 1 | - |
| dbc:Politics | 118 | 26 |
| dbc:Science | **1529** | **2391** |
| dbc:Society | **979** | **1650** |
| dbc:Technology | **987** | 948 |

As can be noticed in Table 10, the second dataset does not contain `dbc:People` entities. Therefore, the lists representing the datasets do not always have all the 23 top-level categories, i.e., their lists of categories are not always conjoint.

The *entity similarity processor* chooses different similarity measures to compare the ranked lists and generate a similarity score. In the example shown in Table 10, if the similarity measure chosen is the *cosine distance*, the score is 0.784. Choosing *RBO* as the similarity measure (see Section 3), the score is 0.887 (with $p = 0.98$) and 0.940 (with $p = 0.99$). In turn, the *Jaccard distance* gives 0.956 as the similarity score.

Although the Jaccard distance gives the highest similarity score (since the lists have 22 out of the 23 categories in common), the two datasets appear in different communities in the LOD Cloud (considered as the ground truth for the experiment): the *eu-agencies-bodies* is in the *Publication* community and the *rkb-explorer-citeseer* is in the

---

[5] http://linkeddatacatalog.dws.informatik.uni-mannheim.de/sk/dataset/eu-agencies-bodies
[6] http://linkeddatacatalog.dws.informatik.uni-mannheim.de/sk/dataset/rkb-explorer-citeseer
[7] http://citeseer.ist.psu.edu/index

*Government* community. In fact, the relevance of the categories representing the datasets is very different. In these cases, the Jaccard distance is not a reasonable option, since it is not unlikely that two datasets have several categories in common (in total, there are 23 categories), which makes the Jaccard similarity score usually high.

*A ground truth for the dataset's domain*

The Linked Open Data cloud[8] diagram describes datasets that have been published as Linked Data based on metadata collected and curated by contributors to the Data Hub.

We constructed the ground truth from a fragment of the August 2014 version diagram. Each circle represents a dataset and the circle size indicates the number of edges connected to each dataset. The circle color indicates the dataset community. In this version of the diagram, there is a total of 568 datasets, classified into 9 communities: Government (136 datasets), Publications (133 datasets), Social Networking (89 datasets), Life Sciences (63 datasets), User-generated content (42 datasets), Cross-domain (40 datasets), Geographic (24 datasets), Media (21 datasets) and Linguistics (20 datasets).

We extracted the 568 datasets from the 2014 LOD Cloud (see Figure 7), parsing the SVG file to read the circles color to identify the datasets communities (see Figure 9). Then, we selected 7 meaningful communities, discarding the Cross-Domain and the Linguistic communities, since they mix different dataset domains.

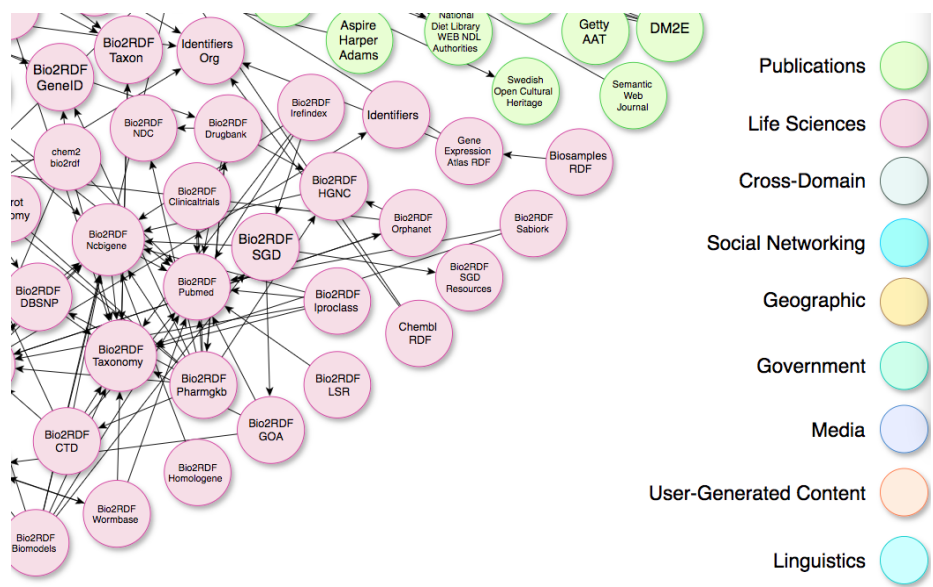

**Figure 9.** LOD cloud diagram fragment

*Evaluating the results*

For the final evaluation, we considered only the 125 datasets present in both the LOD ground truth (with 568 datasets) and in our datasets (with 390 datasets) (see Figure 7).

We consider as baselines two well-known similarity measures: Jaccard distance and Cosine distance. We compared the datasets with each other using the three different similarity measures: Jaccard distance, Cosine distance (the two baselines) and RBO. We generated a distance matrix representing the distance between all dataset pairs. From such distances, we generated 7 clusters, using the Hierarchical Agglomerative algorithm with different linkage criteria: Single, Complete, Ward and Average. To evaluate the proposed entity similarity metric for the dataset's domain, we clustered the datasets using the proposed entity similarity metric and compared the clusters thus obtained with the ground truth communities (or clusters). We assumed that the most similar datasets should belong to the same category.

---
[8] http://lod-cloud.net/

As adopted by Garcia et al. [32], we chose the hierarchical agglomerative clustering algorithm [33], which starts with each dataset as a single cluster and then merges pairs of clusters, using similarity measures, until achieving the desired number of clusters.

To evaluate the clustering performance, we used the Adjusted Rand Index (ARI) (see Section 3). Table 11 shows the ARI values by considering two types of parameters: (a) the similarity measure used to compare the datasets before clustering them (represented by the lines of Table 11); and (b) the clustering linkage metric (see Section 3) used to merge the clusters (represented by the columns of Table 11).

**Table 11:** Adjusted Rand Index of the clustering algorithms

|                   | Single | Complete | Average | Ward  |
|-------------------|--------|----------|---------|-------|
| **Jaccard**       | 0.018  | 0.170    | 0.242   | 0.142 |
| **Cosine**        | 0.161  | 0.277    | 0.284   | 0.267 |
| **RBO, *p* = 0.98** | 0.008  | 0.281    | **0.302** | **0.298** |
| **RBO, *p* = 0.99** | 0.008  | 0.205    | 0.149   | 0.273 |

The worst performance, measured by the ARI index, was obtained using the Jaccard distance. This was expected since Jaccard considers only the presence or the absence of an item in the lists. The cosine distance performed better than Jaccard since it considers the frequency of the categories that describe the datasets. The best performances (0.302 and 0.298) were obtained using *RBO* (with $p$ = 0.98) as a similarity measure, and the *Average* and *Ward* as clustering linkage metrics, respectively.

Figure 10 shows the confusion matrix (a confusion matrix compares two classification models) for the best performing RBO case (with ARI index 0.302, for p = 0.98 and Average as the clustering linkage metric) as compared with the ground truth classification. The lines of the confusion matrix correspond to the communities (or clusters) of the ground truth. They represent, from 0 to 6, respectively, Government, Geographic, Publications, Life Sciences, User Generated Content, Social networking, and Media. The columns, in turn, represent the clusters found in the experiment, using the best performing RBO case.

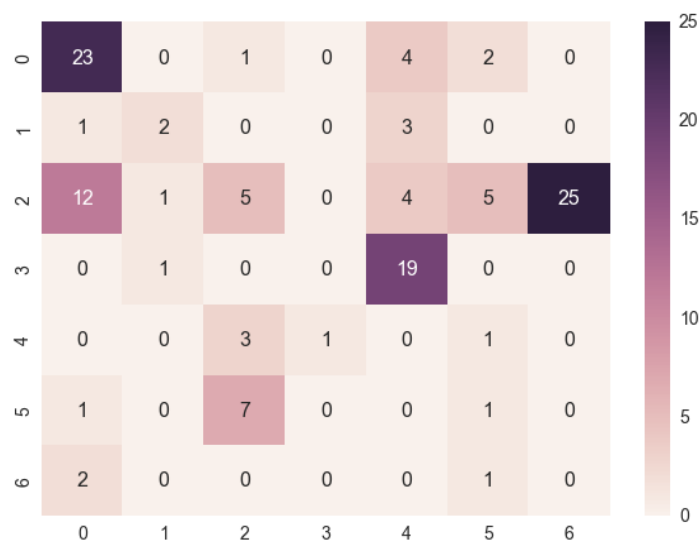

**Figure 10.** Confusion matrix for the best performance case

Analyzing the quality of the generated clusters, we note that Government (line 0), Publications (line 2), Life Sciences (line 3) and Social networking (line 5) were clearly recognized as communities. For the Government community, from a total of 30 datasets in the ground truth, 23 were assigned to the same cluster. For the Publications community (line 2), 25 out of 52 were assigned to the same cluster. For the Life Sciences (line 3), 19 from 20 datasets were in the same clusters. The other communities (Geographic, User Generated Content, Social networking and Media) were not recognized. A possible reason for this lies in their low density in the ground truth. In fact, the Social Networking community has 9 datasets, the Geographic community has 6 datasets, the User Generated content community has 5 datasets, and the Media community has only 3 datasets.

This experiment demonstrates that the best performing algorithm is that which consider the entity features as ranked lists, in our case, the RBO metric.

### 5.1.3 Comparing DBLP Computer Science conferences

*Computer Science Conferences in DBLP*

The DBLP repository[9] stores Computer Science bibliographic data for more than 4,500 conferences and 1,500 journals. DBPL is a joint service of the University of Trier and the Schloss Dagstuhl. Table 12 shows DBLP statistics in August 2017. In this experiment, we extracted computer science conferences from the DBLP repository to instantiate the SELEcTor framework.

**Table 12.** DBLP statistics in August 2017

| Entity type | Number of entities |
|---|---|
| Publications | 3,859,721 |
| Authors | 1,946,939 |
| Conferences | 5,163 |
| Journals | 1,544 |

---
[9] http://dblp.uni-trier.de/

*Extracting and ranking features*

We considered the keywords extracted from the papers published in a conference as features to describe the conference. We extracted the stem-words from the keywords, in order to cope with different variations of the same root term. For instance, the *retriev* stem-word matches both with the *retrieval* and with the *retrieving* keywords; analogously, the *relev* stem-word matches both with *relevant* and with the *relevance* keywords. This strategy is detailed in (García et al., 2017). Table 13 shows the top 16 stem-words (out of 1,847 stem-words) extracted from papers from the SIGIR conference[10], the International ACM Conference on Research and Development in Information Retrieval.

**Table 13: SIGIR top stem-words**

| stem-word | frequency |
|---|---|
| retriev | 808 |
| search | 600 |
| inform | 551 |
| queri | 475 |
| model | 467 |
| web | 317 |
| base | 261 |
| document | 255 |
| text | 244 |
| evalu | 239 |
| rank | 232 |
| languag | 220 |
| relev | 216 |
| learn | 203 |
| user | 189 |
| cluster | 179 |

*A ground truth for the conference's domain*

We chose as ground truth for the experiment the list of academic Computer Science conferences defined in Wikipedia[11], with 248 conferences grouped into 13 groups. Although the 13 groups are subdivided into smaller groups, we considered only the 13 more general groups available, listed in Table 14, together with the number of conferences of each group.

**Table 14:** The Wikipedia groups of conferences

| Group | # conferences |
|---|---|
| Artificial intelligence | 38 |
| Computer networking | 35 |
| Languages and software | 27 |
| Algorithms and theory | 27 |
| Computer architecture | 25 |
| Concurrent, distributed and parallel computing | 24 |
| Data Management | 21 |
| Security and privacy | 14 |
| Computer graphics | 9 |
| Human-computer interaction | 9 |
| Operating systems | 8 |
| Education | 6 |
| Computational biology | 5 |

---

[10] http://dblp.uni-trier.de/db/conf/sigir/
[11] https://en.wikipedia.org/wiki/List_of_computer_science_conferences

*Computing entity similarity*

Our strategy to compare the conferences was analogous to the strategy adopted for the dataset experiment (see Section 5.1.2).

First, we compared all 248 conferences using similarity measures to generate a similarity matrix, with 61,256 cells, from which 30,380 cells (the lower triangular part of the matrix) are filled with the similarity between the pairs of distinct conferences. Then, we clustered the conferences using the hierarchical agglomerative clustering algorithm.

*Evaluating the results*

We consider again as baselines the well-known similarity measures: Jaccard distance and Cosine distance. To evaluate the clustering process, we again adopted the Adjusted Rand Index (ARI). Table 15 shows the ARI values for two parameters: (a) the similarity measure used to compare the conferences before clustering them; and (b) the clustering linkage metric used to merge the clusters when executing the hierarchical agglomerative clustering.

Table 15: Adjusted Rand Index (ARI) for the clustering algorithms comparing conferences

|  | Single | Complete | Average | Ward |
|---|---|---|---|---|
| **Jaccard** | 0.343 | 0.599 | 0.612 | 0.586 |
| **Cosine** | 0.343 | 0.589 | 0.630 | 0.713 |
| **RBO, $p = 0.97$** | 0.464 | 0.598 | **0.794** | 0.602 |
| **RBO, $p = 0.98$** | 0.562 | 0.661 | 0.742 | 0.727 |
| **RBO, $p = 0.99$** | 0.361 | 0.670 | 0.754 | 0.727 |

The worst performance was again obtained using the Jaccard distance. The cosine distance had better results when combined with Ward as linkage criteria (ARI=0.713). Note that, in general, the RBO had the best performances. The best overall performance, with ARI=0.794, was obtained using RBO as a similarity measure, with *p*=0.97, and *Average* as clustering linkage metric.

We performed the experiments using Python with Jupyter in a Macbook air 1,6 GHz Intel Core i5 4 GB 1600MHz DDR3. Using the Jaccard distance, it took 27 seconds to construct the 248×248 similarity matrix (for the 248 conferences). Using RBO, it took around 210 seconds. Using the Cosine distance, it took around 25 hours. By contrast, regarding the previous two experiments, since the number of entities to be compared was considerably smaller (12 museums and 125 datasets), the computational cost of the similarity measures was negligible.

Figure 11 shows the confusion matrix for the best performance case (ARI=0.794). The lines correspond to the ground truth clusters and the columns refer to the clusters generated by the best performing case.

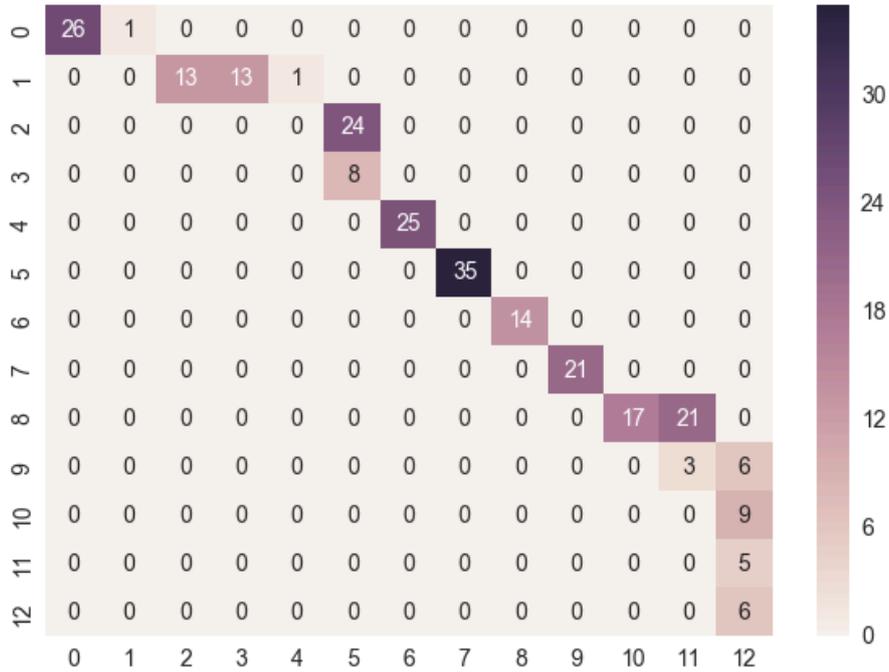

**Figure 11:** Confusion matrix for the best performance case

By analyzing the quality of the generated clusters, we notice that four conference groups were entirely identified by the clustering process: The *Computer architecture* group (line 4), with 25 conferences, the *Computer networking* group (line 5), with 35 conferences, the *Security and privacy* group (line 6), with 14 conferences, and *Data Management,* with 21 conferences (line 7). The *Algorithms and theory* group (line 0) had 26 out of its 27 conferences assigned to cluster 0 and one conference to cluster 1. From the 27 conferences of the *Languages and software* group (line 1), 13 were assigned to cluster 2, 13 to cluster 3, and one conference to cluster 4. All 24 conferences of the *Concurrent, distributed and parallel computing* group (line 2) went to cluster 4, together with 8 other conferences from the *Operating systems* group (line 3). The biggest group, *Artificial intelligence* (line 8, 38 conferences), had 17 conferences in cluster 10, and 21 in cluster 11. The *Computer graphics* group (line 9) had 3 conferences assigned to cluster 11 and the other 6 to cluster 12. Finally, the last three groups, *Human-computer interaction*, with 9 conferences (line 10), *Computational Biology*, with 5 conferences (line 11), and *Education*, with 6 conferences (line 12), were merged into only one cluster, cluster 12.

Therefore, this experiment also demonstrates that the best performing algorithm is that which consider the entity features as ranked lists, in our case, the RBO metric. Also, it demonstrates that the cosine distance would be a reasonable option, only if the number of entities is fairly small.

# 6    Conclusions

This paper addressed the problem of estimating semantic entity similarity using entity features available as Linked Data. The key idea was to exploit ranked lists of features, extracted from Linked Data sources, as a representation of the entities to be compared. The semantic similarity between the two entities was then estimated by comparing their ranked lists of features. We argued that, by ranking the features based on their relevance, we improve the accuracy of the similarity computation.

We generalized an approach to estimate entity similarity using ranked features extracted from Linked Data sources, proposed in our previous work, called SELEcTor. Then we proved the generality of the SELEcTor framework, by instantiating it, in three different domains, and by carrying out detailed experiments. In the first experiment, we compared museums represented in DBpedia. We found that the art movements of the museums' artworks are high-quality features. In the second experiment, we compared datasets represented in a Linked Data repository, using their Wikipedia top-level categories as features. Finally, in the last experiment, we compared computer science conferences, also provided as Linked Data in the DBLP repository, using the keywords extracted from their publications as features. We achieved better results than chosen baselines in all experiments.

As for future work, we intend to address other domains using our approach to compute entity similarity. For example, an interesting domain refers to university departments or institutes. When comparing two computer science departments (entities), say, one could profitably use the approach we presented in this article by first generating ranked lists of published works (features) in conference or journal, and then comparing the ranked lists to compute the similarity between the departments. By analogy with the museum's scenario, the departments would be analogous to the museums, the published works – journal or conference papers – to the museum artworks, and the authors – professors, researches, and students – to the artists.

## Acknowledgments

This work was partly funded by CNPq under grants 303332/2013-1 and 442338/2014-7, by FAPERJ under grant E-26/201.337/2014 and by H2020-MSCA-RISE project MASTER GA 777695.